\documentclass[aps,amssymb,showkeys]{revtex4}
\usepackage{amsmath}
\usepackage[dvips, bookmarks, colorlinks=true, plainpages = false, citecolor = red, urlcolor = blue, filecolor = blue]{hyperref}
\def\be{\begin{equation}}
\def\ee{\end{equation}}
\def\nn{\nonumber} 
\newcommand{\je}[2]{\mathrm{#1}\left(#2 K'|m\right)}
\def\p{\partial} 
\def\G{\Gamma}
\newcommand{\drv}[2]{\partial_{#2} #1}
\newcommand{\Drv}[3]{\partial_{#2}^{#3} #1}
\newcommand\e{\varepsilon} 
\def\k{\kappa}
\def\y{\psi}
\def\Real{\mathrm{Re}}
\def\Imag{\mathrm{Im}}
\def\I{\mathrm{i}}

\begin{document}

\title{Complete conformal field theory solution of a chiral six--point correlation function}

\author{Jacob J. H. Simmons}
\email{simmonsj@uchicago.edu}
\affiliation{Rudolf Peierls Centre for Theoretical Physics, 1 Keble Road, Oxford OX1 3NP, UK}
\affiliation{James Franck Institute, 929 E 57th Street, Chicago, IL 60637, USA}

 \author{Peter Kleban}
\email{kleban@maine.edu}
\affiliation{LASST and Department of Physics \& Astronomy,
University of Maine, Orono, ME 04469, USA}


%%%%%%%%%%%%%%%%%%%%
%                    ABSTRACT                        %
%%%%%%%%%%%%%%%%%%%%
\begin{abstract}
Using conformal field theory, we perform a complete analysis of the chiral six-point correlation function
\be \nonumber
C(z)=\langle \phi_{1,2}\phi_{1,2} \Phi_{1/2,0}(z, \bar z) \phi_{1,2}\phi_{1,2} \rangle \; ,
\ee
 with the four $\phi_{1,2}$ operators at the corners of an arbitrary rectangle, and the point $z = x+iy$ in the interior. We calculate this for arbitrary central charge (equivalently, SLE parameter $\k > 0$).  $C$ is of physical interest because for percolation ($\k = 6$) and many other two-dimensional critical points, it specifies the density at $z$ of critical clusters conditioned to touch either or both  vertical ends of the rectangle, with these ends `wired', i.e.\ constrained to be in a single cluster, and the horizontal ends free.  

The correlation function may be written as the product of an algebraic prefactor $f$ and a conformal block $G$, where  $f = f(x,y,m)$, with $m$ a cross-ratio  specified by the corners ($m$ determines the aspect ratio of the rectangle).  By appropriate choice of $f$ and using coordinates that respect the symmetry of the problem, the  conformal block $G$ is found to be independent of either $y$ or $x$, and given by an Appell function.

\end{abstract}
\maketitle

%%%%%%%%%%%%%%%%%%%%
%                INTRODUCTION                   %
%%%%%%%%%%%%%%%%%%%%
\section{Introduction}

The methods of conformal field theory (CFT)  \cite{BPZ84, BYB} allow calculation of the correlation functions of a variety of operators, which may, in many cases, be interpreted as physical quantities in the continuum limit of two dimensional critical systems.  As the number of operators grows, the calculation becomes progressively more difficult, however, and in practice there are very few results for correlators with more than four operators.

In this paper we present a full calculation of the six-point correlator
\be \label{cf1}
C(z)=\langle \phi_{1,2}^c(0)\phi_{1,2}^c(\I) \Phi_{1/2,0}(z, \bar z) \phi_{1,2}^c(R)\phi_{1,2}^c(R+\I) \rangle_{\cal R}\; ,
\ee 
in the rectangular geometry  $\mathcal{R}:= \{z=x + \I y\in\mathbb{C}\, |\, 0<x<R, 0<y<1\}$.  The conformal dimensions used through out this article and the associated central charge are  
\begin{align} \label{hcdims}
h_{1,2} &= \frac{6-\k}{2\k} &
h_{1/2,0} = \bar h_{1/2,0}&=\frac{(8-\k)(3\k-8)}{64 \k} \\ \nn
h_{1,3}&= \frac{8-\k}{\k}  &
c &= \frac{(3 \k -8)(6-\k)}{2 \k} \; ,
\end{align}
where $\k$ is the Schramm-Loewner Evolution (SLE) parameter.   Because of their positions the $\phi_{1,2}^c$ corner operators have an effective dimension $h^c_{1,2}=2h_{1,2}$, i.e.~twice the usual value (see subsection \ref{Cff} for more details on this point).  

$C(z)$ is of interest because it may be used in a variety of physical models to determine the density of clusters attached to one, or both, of two distinct boundary intervals, when these intervals are each `wired', i.e.\ constrained to belong to a single cluster.  A recent paper \cite{SimmonsZiffKleban08} presents a calculation of  $C$ for percolation ($\k = 6$) in a semi-infinite rectangle, and also in an arbitrary rectangle by assuming a certain $y$--independence specific to the rectangle.

Here, we  calculate $C$ completely and without assumptions,  for arbitrary $\k >0$.  This is done by solving the differential equations that $C$ satisfies.  The main new step  is a certain choice of coordinates.  This choice allows us to derive the curious $y$--independence mentioned, to reduce the number of variables from three to two, and write the solutions explicitly in terms of Appell functions.  The application of these results to a variety of physical systems is considered in \cite{SimmonsKlebanFloresZiff11}.

To begin, we express $C$ in the form
\be \label{CfGeq}
C(z) = f(z) \,F(z) \; ,
\ee 
where $f(z)$ is an appropriately chosen algebraic prefactor. With this choice  it transpires that for a given rectangle, all possible solutions for $F(z)$ can be written in the form $F(z)=G(x)+\widetilde{G}(y)$, where $\widetilde{G}$ is determined by $G$.  Then, using certain elliptic functions of $x$ and $y$ as intermediate variables, we find an algebraic expression for $f$, and algebraic factors times Appell hypergeometric functions for $G$. The full expressions for $f$ and $G$ depend on the aspect ratio of the rectangle as well as $x$ and/or $y$.

The prefactor $f$ is independent of the details of the physical system, while $F$ changes depending on the particular observable associated with $C(z)$.  For a conformal block of (\ref{cf1}), $F(z)$ only depends on $x$ or $y$.   This surprising feature, originally observed numerically (see \cite{SimmonsZiffKleban08}),  indicates the presence of some unknown symmetry.  It also implies that in a given rectangle appropriate ratios of two $C(z)$ with different physical meanings are completely independent of either $x$ or $y$, since the prefactor $f$ cancels out.

In a companion paper \cite{SimmonsKlebanFloresZiff11} we apply these results in several ways.  First, to find the cluster densities for a range of critical $\mathrm{O}(n)$ loop models, in both dense and dilute phases and equivalently, for critical $Q$-state Potts models, probing either FK or spin clusters.   Second, we extend previous results for the factorization of correlations for percolation, described in  \cite{SimmonsZiffKleban08}, to the critical models mentioned.  Finally, for percolation,  the density of horizontal crossing clusters in a rectangle with open boundary conditions on all edges is determined.

In this work,  section \ref{Th} calculates the correlation function (\ref{cf1})  by solving the associated PDEs,  which with proper choice of co-ordinates and prefactor $f$ reduce to the Appell equations. Subsection \ref{cfDEs} gives the PDEs, choses $f$ and coordinates, and presents the solutions for $G$, all of which are single conformal blocks.  The interesting independence of $G$ from one coordinate mentioned (see  (\ref{DE3})) appears here.   Subsection \ref{Cff} computes the form of the correlation function prefactor $f$ in the rectangle, a not completely trivial task.  Section \ref{Conc} contains a summary of our results and some discussion. 

In Appendix \ref{F2Der}, we derive a relation between Appell functions that is useful in simplifying one of our formulas.   Appendix \ref{Ccd} examines conditions for a set of conformal blocks  to have a common $y$--dependence in the rectangle.

%%%%%%%%%%%%%%%%%%%%
%                     THEORY                            %
%%%%%%%%%%%%%%%%%%%%
\section{Theory} \label{Th}
%%%%%%%%%%%%
%      DIFF. EQU.S        %
%%%%%%%%%%%%
\subsection{Solving the differential equations} \label{cfDEs}
In this subsection we determine and solve the differential equations for the correlation function (\ref{cf1}).  The main new step in solving the differential equations is a certain choice of coordinates, given below.  This choice allows us to derive the interesting $y$--independence mentioned, and to write the solutions in terms of Appell functions.
  
To begin, we consider $C$ in the upper half plane $\mathbb{H}:=\{w=u+\I v\, |\, u \in \mathbb{R}, v>0 \}$
\be \label{cf0}
C(w) = \langle \phi_{1,2}(u_1)\phi_{1,2}(u_2) \Phi_{1/2,0}(w, \bar w) \phi_{1,2}(u_3)\phi_{1,2}(u_4) \rangle_{\mathbb{H}} \; ,
\ee
using the methods of boundary CFT, then transform into the rectangle $\mathcal{R}$.   In  $\mathbb{H}$ we can decompose $\Phi_{1/2,0}(w,\bar w)$ into chiral components $\Phi_{1/2,0}(w)\Phi_{1/2,0}(\bar w)$. Then, by conformal symmetry one may write
\be  \label{UHP CI Form}
C(w) = \frac{|w-\bar w|^{h_{1,3}-2 h_{1/2,0}}\left[ u_{31}u_{42}\right]^{\frac{1}{2}h_{1,3}-2h_{1,2}}}{\left| (w-u_1)(w-u_2)(w-u_3)(w-u_4) \right|^{\frac{1}{2}h_{1,3}}}F\left( \frac{(w-u_1)u_{43}}{u_{31}(u_4-w)},\frac{(\bar w-u_1)u_{43}}{u_{31}(u_4-\bar w)},\frac{u_{21} u_{43}}{u_{31}u_{42}} \right)\; , 
\ee 
where $u_{ij}:=u_i-u_j$. In the next section we give our reason for choosing this particular form for $C(w)$.

Using standard CFT methods  \cite{BPZ84} $F$ may be determined via the differential equations arising from  the null state  $\left[3L_{-1}{}^2-2(1+2 h_{1,2})L_{-2} \right]|\phi_{1,2}\rangle$ associated with each $\phi_{1,2}$ in   (\ref{UHP CI Form}).  The presence of $\phi_{1,2}(u_1)$, for instance, means that (\ref{UHP CI Form}) is annihilated by the operator 
\be \label{DO1}
\frac{2h_{1/2,0}\,\Real\left[ (w-u_1)^2 \right]}{|w-u_1|^4}-\frac{2\, \Real \left[ (\bar w-u_1)\p_w \right]}{|w-u_1|^2}+\sum_{j=2}^4\left[ \frac{h_{1,2}}{(u_j-u_1)^2}-\frac{\p_{u_j}}{u_j-u_1}\right]-\frac{3}{2(1+2h_{1,2})} \p_{u_1}{}^2 \; .
\ee

Next we let $\{ u_1,u_2,u_3,u_4\} \to \{0, m, 1, \infty\}$, which means that under the conformal map to the rectangle (\ref{ConfMapping}) $m$ is the cross-ratio $m = (u_{12}u_{34}/u_{13}u_{24})$ of the image points of the corners of the rectangle.  (Note that $m$  differs from  the standard modular lambda parameter, which is $1-m$ here.)  Thus one arrives at a differential equation for $F(w,\bar w, m)$,
\begin{align} \nonumber
0=&\,
\frac{8(6-\k )w \bar w +(8-\k )m(4(2 m-(w+\bar w ))+(8-\k ) m(1-w)(1-\bar w ))}{4\k^2  m^2 w  \bar w }F
\\ \nonumber
&+\frac{(1-m)((8-\k ) m(w+ \bar w )+2w \bar w(\k  m-4))}{2 \k m w  \bar w } \drv{F}{m}
\\  \label{DE1}
&+\frac{(1-w) ((8-\k )w-\k  \bar w +2\k  w \bar w )}{2 \k w \bar w } \drv{F}{w}
+\frac{(1-\bar w ) ((8-\k)\bar w-\k w +2 \k w \bar w )}{2\k w \bar w }\drv{F}{\bar w}
\\ \nonumber
&- (1-m)^2 \drv{\drv{F}{m}}{m}
-2 (1-w) (1-m) \drv{\drv{F}{m}}{w}
-2 (1-m ) (1-\bar w )\drv{\drv{F}{m}}{\bar w}
\\ \nonumber
&-(1-w)^2 \drv{\drv{F}{w}}{w}
-2 (1-w) (1-\bar w ) \drv{\drv{F}{w}}{\bar w}
-(1-\bar w )^2 \drv{\drv{F}{\bar w}}{\bar w}
\; .
\end{align}

We next  transform (\ref{DE1}) %
  into rectangular coordinates via the conformal mapping
\be \label{ConfMapping}
w(z, m)=m\,\mathrm{sn}\left(z\, K'|m\right)^2, \quad
\bar w(\bar z, m)=m\,\mathrm{sn}\left(\bar z\, K'|m\right)^2, 
\ee
where $K' :=K(1-m)$, with $K:=K(m)$  the complete elliptic integral of the first kind, and $\mathrm{sn}(\cdot|m)$ the Jacobi elliptic function with elliptic parameter $m$.  The factor  $K'$  appears because our ${\cal R}$ has fixed height of $1$, which differs from the standard rectangle used to define the Jacobi elliptic functions.  The aspect ratio is given by 
\be \label{Rvsm}
R=\frac{K(m)}{K'(m)} \; ,
\ee
 which is the inverse of the standard elliptic aspect ratio.  Conversely, the aspect ratio $R$ specifies the elliptic parameter $m$  via
\be \label{mvsR}
m=\frac{\vartheta_4{}^4\left(0,e^{- \pi R} \right)} {\vartheta_3{}^4\left(0,e^{- \pi R} \right)}\; .
\ee
The transformation (\ref{ConfMapping}) maps the corners of the rectangle  $\mathcal{R}$, starting at the origin and proceeding counterclockwise,  into the $w$-plane points $(0,0), (m,0), (1,0)$ and $(\infty)$, respectively. %

We now introduce the real coordinates
\be \label{xipsidef}
\xi=\je{sn}{x}^2\; , \qquad \mathrm{and} \qquad \y=\mathrm{sn}\left(y\, K'|1-m\right)^2\; .
\ee
This choice of co-ordinates is a key step, as we will see.  It simplifies the equations, leading to a  solution of the PDEs. In addition, our results are either algebraic or hypergeometric  when written in terms of $\xi$ and $\psi$.

 The conformal transformation now becomes
 \be
w(\xi,\psi,m)=m\frac{\xi (1-(1-m) \y)-(1-\xi)(1-m \xi)\y(1-\y)}{(1-\y+m \xi \y)^2}+i\, m\frac{2\sqrt{\xi(1-\xi)(1-m \xi)\y(1-\y)(1-(1-m)\y)}}{(1-\y+m \xi \y)^2}\; .
\ee

The coordinate $\xi$, to within a factor $m$,  is the half-plane image of the projection $z\mapsto x$ onto the bottom edge of the rectangle, i.e.~$w(x)=m\,\xi$ so that $\xi \in (0,1)$.  The coordinate $\psi$ is slightly more complicated and is determined by first taking the projection $z\mapsto \I y$ on the left end of the rectangle.  The half-plane image of this point is in the interval $(-\infty,0)$, and we define $\psi$ via $w(\I y)=:m \y/(\y-1)$ so that $\y \in (0,1)$.  We will see in the following paragraphs that this choice of coordinates introduces a useful $\xi \leftrightarrow \psi$ symmetry and allows us  to compare directly with results from \cite{SimmonsZiffKleban08}.

Transforming (\ref{DE1}) into these coordinates (with the help of Mathematica) yields
\begin{align} \label{DE2}
0=&\left(8(6-\kappa )(\xi +\psi -\xi  \psi )^2+(8-\kappa )^2(1-m \xi -\psi+m\psi )^2 \right.\\ \nonumber
&\left.\qquad +8(8-\kappa )\left((1-\xi )(1-\psi )-\xi  \psi (1-m \xi )(1-\psi+m\psi)\right)\right)F\\ \nonumber
&+2\kappa  m(1-m)\left(((8-\kappa ) (\xi -\psi )-8 \xi  \psi(1-2 m))(2-\xi -\psi +\xi  \psi )
-(1-2 m) \kappa  \left(\xi ^2-\xi ^2\psi + \psi ^2-\xi  \psi ^2\right)\right)\drv{F}{m}\\ \nonumber
&+4\kappa  \xi (1-\xi )\left(4 \left(1-\psi +m^2 \xi  \psi \right)-(\kappa -2\kappa  \xi +4\xi )(1-\psi +m \psi )^2 
-(\kappa -4) \xi  m(1-\psi +m \psi ) \right) \drv{F}{\xi}\\  \nonumber
&+4\kappa  \psi (1-\psi )\left(4 \left(1-\xi +(1-m)^2 \xi  \psi \right)-(\kappa -2\kappa  \psi +4\psi )(1-m \xi )^2
-(\kappa -4) \psi (1-m)(1-m \xi ) \right) \drv{F}{\psi }\\ \nonumber
&-4\k^2\, m^2(1-m)^2(\xi +\psi -\xi  \psi )^2 \Drv{F}{m}{2}
-8\k^2\, m(1-m) \xi (1-\xi )(1-(1-m) \psi )(\xi +\psi -\xi  \psi ) \drv{\drv{F}{\xi}}{m}\\ \nonumber
&+8\k^2\, m(1-m) (1-m \xi ) \psi (1-\psi )(\xi +\psi -\xi  \psi )\drv{\drv{F}{\y}}{m}
-4\k^2\, \xi ^2(1-\xi )^2(1-(1-m) \psi )^2 \Drv{F}{\xi}{2}\\ \nonumber 
&+8\k^2\, \xi (1-\xi )(1-m \xi )\psi (1-\psi )(1-(1-m) \psi ) \drv{\drv{F}{\xi}}{\y}
-4\k^2\, (1-m \xi )^2\psi ^2(1-\psi )^2 \Drv{F}{\y}{2}\; .
\end{align}

Three additional equations are derived by cyclicly permuting the indices on the $u$ variables in (\ref{DO1}) and following the steps above.  These three additional equations can also be obtained from (\ref{DE2}) by the conformal symmetries of rectangles with arbitrary $R$ and fixed height $1$: reflection about $x=R/2$, reflection about $y=1/2$, and reflection over $x=y$ with a concurrent scaling by $1/R$ in order to preserve the height.  The last of these may be implemented by a change of aspect ratio $R\to1/R$ (i.e.\ $m \to 1-m$), exchanging $x$ and $y$, and then  scaling by a factor of $1/R$, so that, for example, $\xi =\mathrm{sn}\left(x\, K'|m\right)^2  \to \mathrm{sn}\left(y\, K (K'/K)|1-m\right)^2 =\y$.  The third symmetry operation introduces a conformal covariance factor due to the scaling, but this is absorbed into the prefactor and does not effect $F$.  The three symmetry operations translate, respectively, into 
\be  \label{Sym}
( \xi, \y, m ) \to  \left( \frac{1-\xi}{1-m\, \xi}, \y, m\right), \left(\xi, \frac{1-\y}{1-(1-m) \y}, m \right), \left(\y, \xi,1-m \right) \; .
\ee

Now comes a central mathematical result.  There is a linear combination of the four differential equations giving 
\be \label{DE3}
\drv{\drv{F}{\xi}}{\y}(\xi, \y, m)=0\;.
\ee
(Since $m$ is a variable here we have altered the usual Jacobian notation and write e.g.~$F(\xi, \y, m)$ in place of $F(\xi, \y|m)$.)
Thus all solutions must be of the form 
\be \label{SepForm}
F(\xi, \y, m)=G(\xi, m)+\widetilde{G}(\y, m)\; .
\ee
(An $F$ depending on $m$ alone is not possible, as follows from (\ref{DE2})).

Now, for a given aspect ratio, $\xi$ depends only on $x$ and $\psi$ depends only on $y$.  As is spelled out in detail in \cite{SimmonsKlebanFloresZiff11}, (\ref{DE3})  then implies that the ratio of correlation functions investigated in  \cite{SimmonsZiffKleban08} for percolation is independent of $y$ (the horizontal coordinate) in the rectangle.  This peculiar symmetry was, as noted in \cite{SimmonsZiffKleban08}, first observed via simulations.  Using  (\ref{DE3}) shows that this symmetry is exact in the continuum limit, and holds for  many critical systems.

Because of the symmetry $\{ \xi \leftrightarrow \y , m   \leftrightarrow  1-m \}$, which follows from (\ref{xipsidef}), it is sufficient to find the solutions of $F(\xi,\y,m)=G(\xi,m)$.  The full solution set can then be obtained by letting $\widetilde{G}(\y,m)=G(\y,1-m)$ in (\ref{SepForm}).  

Inserting $G(\xi,m)$ into the differential equations and taking linear combinations allows us to cancel out all $\y$ dependence and arrive at the three equations
\begin{align} \nonumber
 0 =&\left((8-\kappa )\left(8(1-\xi )+(8-\kappa )(1-m \xi )^2\right)+8(6-\kappa ) \xi ^2\right) G
 -4 \kappa  \xi  m(1-m)(\kappa  (1-m \xi )-4(2- \xi )) \drv{G}{m}\\ \label{DE4}
& +4 \kappa  \xi (1-\xi )(4 \xi -(\kappa -4)(1-2\xi +m \xi ))\drv{G}{\xi}
 -4 \kappa ^2 \xi ^2 m^2(1-m)^2\Drv{G}{m}{2}\\ \nonumber
& -8 \kappa ^2\xi ^2(1-\xi )m(1-m) \drv{\drv{G}{\xi}}{m}
 -4 \kappa ^2 \xi ^2(1-\xi )^2 \Drv{G}{\xi}{2} \; ,\\ \nonumber
0=&\left((8-\kappa )\left(4\left(1-m \xi ^2\right)+(8-\kappa )(1-m)(1-m \xi )\right)-8(6-\kappa ) \xi (1-\xi )\right) G\\   \label{DE5}
&+2\kappa  m (1-m)\left(8-\kappa +8 \xi (1-2 m)-8 \xi ^2+8 m \xi ^2+m \kappa  \xi ^2\right) \drv{G}{m}\\ \nonumber
&+2 \kappa  \xi  (1-\xi )(4(1-\xi )+(1-m)(-2\kappa -4 \xi +4 \kappa  \xi +(8-\kappa ) m \xi ))\drv{G}{\xi}\\ \nonumber
&+4\kappa ^2\xi (1-\xi )m^2 (1-m)^2 \Drv{G}{m}{2}
+4 \kappa ^2\xi (1-\xi )m(1-m) (1-2\xi +m \xi ) \drv{\drv{G}{\xi}}{m}
-4\kappa ^2\xi ^2(1-\xi )^2 (1-m) \Drv{G}{\xi}{2} \; , {\rm \; and} \\ \nonumber
0=&\left((8-\kappa )\left(8\xi (1-m)(1-m \xi )+(8-\kappa )(1-m)^2\right)+8(6-\kappa )(1-\xi )^2\right) G\\ \label{DE6}
&+4\kappa  (1-\xi ) m(1-m)(4+4\xi (1-2m)-(1-m) \kappa ) \drv{G}{m}
-4 \kappa  \xi (1-\xi )(1-m)^2\text{  }(4 \xi +\kappa  (1-2 \xi )) \drv{G}{\xi}\\ \nonumber
&-4 \kappa ^2(1-\xi )^2m^2(1-m)^2 \Drv{G}{m}{2}
+8 \kappa ^2\xi (1-\xi )^2 m(1-m)^2\drv{\drv{G}{\xi}}{m}
-4 \kappa ^2\xi ^2(1-\xi )^2(1-m)^2 \Drv{G}{\xi}{2} \; .
\end{align}
 
In \cite{SimmonsZiffKleban08} we calculated (\ref{cf1}), but only for the case of percolation  ($\k=6$), and in the limit where $z$ goes to the bottom edge of the rectangle.  Here, because the $\y$ dependence  is entirely contained in the prefactor of (\ref{UHP CI Form}) we expect that the  solution space of (\ref{DE4})--(\ref{DE6}) for $\k=6$ should contain the functions calculated in \cite{SimmonsZiffKleban08}, up to differences in the prefactor.  Guided by this, and with a little algebra, we find that making the substitution
\be
G(\xi,m)=\frac{(1-m)^{h_{1,3}-2h_{1,2}}}{m^{2h_{1,2}}\left[\xi (1-\xi )(1-m \xi )\right]^{\frac{1}{2}h_{1,3}}}H(m, m\, \xi)=\frac{(1-m)^{2/\k }}{m^{(6-\k )/\k }\left[\xi (1-\xi )(1-m \xi )\right]^{(8-\k )/(2\k )}}H(m,m\, \xi)\; ,
\ee
and  taking an appropriate linear combination of the resulting equations gives the standard form of Appell's hypergeometric differential equations
\begin{align}\nonumber
0&=\frac{2(8-\k)(\k-4)}{\k^2} H(s,t)+\frac{2(\k-4)-4(\k-5)t}{\k}  \drv{H}{t}+t(1-t) \Drv{H}{t}{2}+\frac{2(8-\k)}{\k} s \drv{H}{s}+s(1-t) \drv{\drv{H}{s}}{t}\; ,\\  \label{AFDE}
0&=-\frac{4(\k-4)}{\k^2} H(s,t)-\frac{4}{\k} t \drv{H}{t}+\frac{2(\k-4)-2\k\, s}{\k}\drv{H}{s}+t(1-s) \drv{\drv{H}{s}}{t}+s(1-s) \Drv{H}{s}{2}\; ,\quad \mathrm{and}\\ \nonumber
0&=-\frac{4}{\k} \drv{H}{t}-\frac{2(8-\k)}{\k} \drv{H}{s}+(t-s) \drv{\drv{H}{s}}{t}\; ,
\end{align}
with $s=m$ and $t= m\, \xi$. Equations (\ref{AFDE}) have a three dimensional solution space.  Among the solutions are five convergent Frobenius series \cite{ApKdF}: 
\begin{align} \nonumber
H_{\mathrm{I}}(s,t)&=s^{-4/\k }t^{12/\k-1}F_1\left(1-\frac{4}{\k };\frac{4}{\k },\frac{4}{\k };\frac{12}{\k }\bigg|\frac{t}{s} ,t \right)\; ,\\ \nonumber
H_{\mathrm{II}}(s,t)&=F_1\left(1-\frac{4}{\k};\frac{4}{\k},2-\frac{16}{\k};2-\frac{8}{\k}\bigg|1-s,1-t \right)\; ,\\ \nonumber
H_{\mathrm{III}}(s,t)&=F_1\left(1-\frac{4}{\k};\frac{4}{\k},2-\frac{16}{\k};2-\frac{8}{\k}\bigg| s,t \right)\; ,\\ \nonumber
H_{\mathrm{IV}}(s,t)&=\frac{(1-t)^{16/\k-2}}{(1-s)^{8/\k-1}}F_1\left(1-\frac{4}{\k};\frac{4}{\k},2-\frac{16}{\k};2-\frac{8}{\k}\bigg|1-s,\frac{1-s}{1-t }\right)\; ,\\ \nonumber
H_{\mathrm{V}}(s,t)&=\frac{(s-t)^{12/\k-1}(1-t)^{4/\k-1}}{s^{4/\k }(1-s)^{8/\k-1}}F_1\left(1-\frac{4}{\k};\frac{4}{\k},\frac{4}{\k};\frac{12}{\k} \bigg |\frac{s-t}{1-t},\frac{s-t }{s(1-t)}\right)\; ,
\end{align}
where 
$$
F_1\left(a;b_1,b_2;c|z_1,z_2\right):=\sum_{i,j=0}^\infty \frac{(a)_{i+j}(b_1)_i(b_2)_j z_1{}^iz_2{}^j}{i!\, j!\, (c)_{i+j}}
$$
with the Pochhammer symbol $(z)_n=\Gamma(z+n)/\Gamma(z)$, is he first of the Appell hypergeometric functions.

This set of Appell functions allows us to write five solutions for $G(\xi,m)$, valid for all $\k > 0$, as
\begin{align} \label{GI}
G_{\mathrm{I}}(\xi,m)&=\frac{\G(2-8/\k)\G(16/\k -1)}{\G(12/\k)\G(1-4/\k)}\frac{\left[m(1-m)\right]^{2/\k}\xi ^{8/\k-1/2}}{\left[(1-\xi)(1-m \xi)\right]^{4/\k-1/2}}F_1\left(1-\frac{4}{\k };\frac{4}{\k },\frac{4}{\k };\frac{12}{\k }\bigg| \xi ,m \xi \right) \; , \\ \label{GII}
G_{\mathrm{II}}(\xi,m)&=\frac{(1-m)^{2/\k}}{m^{6/\k-1}\left[\xi (1-\xi )(1-m \xi )\right]^{4/\k-1/2}}F_1\left(1-\frac{4}{\k};\frac{4}{\k},2-\frac{16}{\k};2-\frac{8}{\k}\bigg|1-m,1-m \xi \right) \; , \\ \label{GIII}
G_{\mathrm{III}}(\xi,m)&=\frac{(1-m)^{2/\k }}{m^{6/\k-1 }\left[\xi (1-\xi )(1-m \xi )\right]^{4/\k-1/2}}F_1\left(1-\frac{4}{\k};\frac{4}{\k},2-\frac{16}{\k};2-\frac{8}{\k}\bigg| m,m \xi \right) \; , \\ \label{GIV}
G_{\mathrm{IV}}(\xi,m)&=\frac{(1-m \xi )^{12/\k-3/2}}{\left[m(1-m)\right]^{6/\k-1}\left[\xi (1-\xi )\right]^{4/\k-1/2}}F_1\left(1-\frac{4}{\k};\frac{4}{\k},2-\frac{16}{\k};2-\frac{8}{\k}\bigg|1-m,\frac{1-m}{1-m \xi }\right)\; , \quad {\rm and}\\ \nonumber
G_{\mathrm{V}}(\xi,m)&=\frac{\G(2-8/\k)\G(16/\k-1)}{\G(12/\k)\G(1-4/\k)}\\  \label{GV}
&\hspace{3em} \times\frac{m^{2/\k}(1-\xi )^{8/\k-1/2}}{(1-m)^{6/\k-1}\xi ^{4/\k-1/2}(1-m \xi )^{1/2}}F_1\left(1-\frac{4}{\k};\frac{4}{\k},\frac{4}{\k};\frac{12}{\k} \bigg |\frac{m(1-\xi )}{1-m \xi },\frac{1-\xi }{1-m \xi }\right) .
\end{align}
For our ranges of $\xi$ and $m$ values, (\ref{GI})--(\ref{GV}) exhaust the convergent Frobenius  series solutions to the differential equations that can be expressed with a single $F_1$.  We can also find  five other convergent Frobenius series solutions that can be expressed with a single Appell function of the second type,
\be
F_2\left(a;b_1,b_2;c_1,c_2 | z_1,z_2\right):=\sum_{i,j=0}^\infty \frac{(a)_{i+j}(b_1)_i(b_2)_j z_1{}^iz_2{}^j}{i!\, j!\, (c_1)_i (c_2)_j}\; .
\ee
With the definition
\be \label{loopeq}
n=-2 \cos\left(4 \pi/\k\right) \; ,
\ee 
(note that $n$ is  the parameter of the $\mathrm{O}(n)$ loop models) we have
\begin{align} 
\label{GVI}
G_{\rm VI}(\xi,m):\hspace{-.2em}&=G_{\rm II}-n G_{\rm I} =  G_{\rm IV}-nG_{\rm V} \\ \nn
&=\frac{\G(2-8/\k)\G(16/\k-1)\G(4/\k)}{\G(1-4/\k)\G(8/\k)^2}\\ \nn
&\hspace{3em}\times [m(1-m)]^{2/\k}\left[\frac{\xi(1-\xi)}{1-m \xi}\right]^{8/\k-1/2}F_2\left(\frac{16}{\k}-1;\frac{4}{\k },\frac{4}{\k };\frac{8}{\k },\frac{8}{\k }\bigg| 1-\xi,\frac{\xi(1-m)}{1-m \xi} \right)\; , \\
\label{GVII}
G_{\rm VII}(\xi,m):\hspace{-.2em}&=G_{\rm III}-nG_{\rm II}    =   G_{\rm V}+(1-n^2)G_{\rm I}   \\ \nn
&=\frac{\G(4/\k)\G(2-8/\k)}{\G(8/\k)\G(2-12/\k)}\frac{[m(1-m)]^{2/\k}\xi^{8/k-1/2}}{[(1-\xi)(1-m\,\xi)]^{4/\k-1/2}}F_2\left(1-\frac{4}{\k};\frac{4}{\k},\frac{4}{\k};\frac{8}{\k},2-\frac{12}{\k} \bigg| m \xi,1-\xi\right)  \; , \\
\label{GVIII}
G_{\rm VIII}(\xi,m):\hspace{-.2em}&= nG_{\rm III}-G_{\rm IV}       = n(2-n^2)G_{\rm I}+(n^2-1)G_{\rm II}   \\ \nn
&=\frac{\G(4/\k)\G(1-8/\k)}{\G(8/\k-1)\G(2-12/\k)}\\ \nn
&\hspace{3em} \times
\frac{(1-m)^{2/k}m^{1-6/\k}}{[\xi(1-\xi)(1-m\,\xi)]^{4/\k-1/2}}F_2\left(1-\frac{4}{\k};\frac{4}{\k},2-\frac{16}{\k};\frac{8}{\k},2-\frac{12}{\k} \bigg| 1-m,m\,\xi\right)  \; ,  \\
\label{GIX}
G_{\rm IX}(\xi,m):\hspace{-.2em}&= nG_{\rm III}-G_{\rm II}      =  n(2-n^2)G_{\rm V}+(n^2-1)G_{\rm IV}  \\ \nn
&=G_{\rm VIII}\left(\frac{1-\xi}{1-m\,\xi},m\right)\;,\hspace{1cm} {\rm and} \hspace{1cm}  \\ 
\label{GX}
G_{\rm X}(\xi,m):\hspace{-.2em}&= G_{\rm III}-nG_{\rm IV}     = G_{\rm I}+(1-n^2)G_{\rm V}   \\ \nn
&=G_{\rm VII}\left(\frac{1-\xi}{1-m\,\xi},m\right)\; .
\end{align}
We derive relations (\ref{GVI})--(\ref{GX})  in appendix \ref{F2Der}.

These Frobenius series solutions are useful, since each one is a single conformal block. It is possible to identify the particular block in each case by examining the leading terms, but this can be done more elegantly using the Coulomb gas formalism (see \cite{SimmonsKlebanFloresZiff11}). 

Since the solution space is  three-dimensional,  two of the $G$s in (\ref{GI})--(\ref{GV}) are not independent.  Using the Coulomb gas formalism one can show that
\begin{align}  \label{linrel1}
2 G_{\mathrm{III}}&=(2-n^2)G_{\mathrm{I}}+nG_{\mathrm{II}}+nG_{\mathrm{IV}}+(2-n^2)G_{\mathrm{V}}\; , \quad \mathrm{and}\\ \label{linrel2}
n G_{\mathrm{I}}+G_{\mathrm{IV}}&=G_{\mathrm{II}}+nG_{\mathrm{V}}\; .
\end{align}
Despite this, it is convenient to consider all five solutions, as well as the alternate forms in (\ref{GVI})--(\ref{GX}) because they have simple interpretations in terms of physical models.

The physical content for $\mathrm{O}(n)$ loop and $Q$-state Potts models is examined in \cite{SimmonsKlebanFloresZiff11}.  The normalizations of these solutions are chosen in part for consistency with the vertex operator formulation used there.  We also show that the particular functions $G_{\rm I}$, $G_{\rm V}$, and $G_{\rm VI}$ form a natural basis of the three dimensional solution space for the critical models mentioned.

The hypergeometric functions in the conformal blocks $G$ simplify for certain $\k$ values, corresponding to various physical models. This is explored in  \cite{SimmonsKlebanFloresZiff11}.

Finally, recall that a second set of conformal block solutions that depend on $\y$ and $m$ follows from $G_i(\y,m) = G_i(\xi,1-m)$.

%%%%%%%%%%%%
%    PREFACTOR        %
%%%%%%%%%%%%
\subsection{Common functional factor and corner operators} \label{Cff}

We next complete the transformation of the correlation function $C(w)$  (\ref{UHP CI Form}) from the upper half plane into the rectangle using (\ref{ConfMapping}), by computing the common functional prefactor  in the rectangle, as a function of $\xi, \psi, m$ (alternatively $x, y, m$) and the parameter $\k$.

We've chosen to write the upper half plane prefactor as
\be\label{UHPprefactor}
f_{\mathbb{H}}(w)=\frac{|w-\bar w|^{h_{1,3}-2 h_{1/2,0}}\left[ u_{31}u_{42}\right]^{\frac{1}{2}h_{1,3}-2h_{1,2}}}{\left| (w-u_1)(w-u_2)(w-u_3)(w-u_4) \right|^{\frac{1}{2}h_{1,3}}}\; .
\ee
This particular form was motivated by the derivation of the analogous correlation function for percolation in the semi-infinite strip, ${\cal S}=\{ {\cal R}\, |\, R \to \infty \}$ \cite{SimmonsZiffKleban08}.  We found that it simplified the analysis to define the prefactor so that in ${\cal S}$ it takes the form $f_{\cal S}(z)=\left[ \langle \Phi_{1/2,0}(z, \bar z) \rangle_{\cal S}\right]^{-11/5}$, where $ \langle \Phi_{1/2,0}(z, \bar z) \rangle_{\cal S}$ is the strip one point function.  For percolation ($\k = 6$), the exponent $-11/5 = (h_{1,3}-2h_{1/2,0})/(-2h_{1/2,0})$ is  the ratio of leading exponents in the bulk-boundary fusion rules when the bulk operator approaches a free versus fixed boundary i.e. $\Phi_{1/2,0}(w, \bar w) \to v^{h_{1,3}-2h_{1/2,0}} \phi_{1,3}(u)$ on a free boundary and $\Phi_{1/2,0}(w, \bar w) \to v^{-2h_{1/2,0}} \mathbf{1}(u)$ on a fixed boundary (see \cite{SimmonsZiffKleban08} or \cite{SimmonsKlebanFloresZiff11} for an explanation of the boundary conditions).  Similarly, we have chosen (\ref{UHPprefactor}) so that  when mapped into the rectangle the prefactor satisfies
 \be \label{prefactor}
f_{\cal R}(z)\propto\left[ \langle \Phi_{1/2,0}(z, \bar z) \rangle_{\mathcal{R}}\right]^{(h_{1,3}-2h_{1/2,0})/(-2h_{1/2,0})}\; .
\ee
The successful analysis of $F(z)$ in the last subsection demonstrates the merit of this choice. In the remainder of this subsection we explicitly determine $f_{\cal R}$.

Because the mapping between the upper half plane and rectangle is singular at the corners  the definition of the boundary condition changing operators placed there needs to be adjusted.
Following  \cite{Cardy84} we use the convention
\be
\phi^c(z_c)=\lim_{\e \to 0} (2 \e)^{-h} \phi(z_c+\e)
\ee
where $\phi^c(z_c)$ defines the rectangular corner operator as a boundary operator $\phi$  approaches a corner $z_c$.  It follows that the conformal weight  $h^c$ of the rectangular corner operator is related to the weight $h$ of the associated boundary operator by $h^c=2 h$.

Thus  we may write the rectangular geometry correlation functions as 
\begin{align}
C(z) &=\langle \phi_{1,2}^c(0)\phi_{1,2}^c(R)\phi_{1,2}^c(\mathrm{i}+R)\phi_{1,2}^c(\mathrm{i})\phi_{1/2,0}(z,\bar z) \rangle_{\mathcal{R}}\\ \label{CeqfG}
&=\lim_{\e_j\to 0} (16 \e_1\e_2\e_3\e_4)^{-h_{1,2}} \langle \phi_{1,2}(\e_1)\phi_{1,2}(R-\e_2)\phi_{1,2}(\mathrm{i}+R-\e_3)\phi_{1,2}(\mathrm{i}+\e_4)\phi_{1/2,0}(z,\bar z) \rangle_{\mathcal{R}}\; .
\end{align}

The prefactor in (\ref{CfGeq}) is given by transforming (\ref{UHPprefactor}) into the rectangular geometry using the transformation law $\phi(z) \to w'(z)^h \phi(w(z))$ and associating the covariance factors with $f_{\cal R}$.  The result is
\be
f_{\cal R}(z)= \lim_{\e_j \to 0} \frac{|w'(z)|^{2 h_{1/2,0}} \prod_{i=1}^4 \left| w'(z(u_j)) \right|^{h_{1,2}}}{(16 \e_1\e_2 \e_3 \e_4)^{h_{1,2}} } f_\mathbb{H}\left(w(z)\right)\; . \ee

To leading order in  the variables $\e_j$ we have 
$$
\{u_1,u_2,u_3,u_4\} = \{m(K' \e_1)^2,\; m-m(1-m)(K' \e_2)^2,\; 1+(1-m)(K' \e_3)^2,\;(K' \e_4)^{-2}\}\; .
$$ 
Using $w'(z)=2 K' \left[w(z)\left(m-w(z)\right)\left(1-w(z)\right) \right]^{1/2}$ and defining
\be \label{fconst}
c(m) = 2^{ h_{1,3}}(K')^{8h_{1,2}+2 h_{1/2,0}}(m(1-m))^{2h_{1,2}} \; ,
\ee
 we find 
\begin{align} \nonumber
f_{\cal R}(z)
&= c(m)\left|\frac{\Imag\, w}{\sqrt{w(m-w)(1-w)}}\right|^{h_{1,3}-2 h_{1/2,0}}\\  \label{fxy}
&= c(m)\left(\frac{\Imag\left[\je{sn}{z}^2\right]}{\left| \je{sn}{z} \je{cn}{z} \je{dn}{z} \right|} \right)^{h_{1,3}-2 h_{1/2,0}}\; . 
\end{align}
We may also write $f_{\cal R}(z)$ as
\begin{align} \label{fxyalt1}
f_{\cal R}(z)
&=  c(m) \left[4\, \je{ds}{2 x}^2+4\, \mathrm{ds}(2 y K'|1-m)^2 \right]^{-\frac{1}{2}h_{1,3}+h_{1/2,0}}\\
&= c(m) \, \left(4 \, (m(1-m))^{1/2} \left[\frac{\displaystyle\vartheta_3{}^2\left(-y\, \pi,e^{-\pi R}\right)}{\displaystyle\vartheta_1{}^2\left(-y\, \pi,e^{-\pi R}\right)} -\frac{\displaystyle\vartheta_3{}^2\left(x\, \pi\, i ,e^{-\pi R}\right)}{\displaystyle\vartheta_1{}^2\left(x\, \pi\, i,e^{-\pi R}\right)}  \right] \right)^{-\frac{1}{2}h_{1,3}+h_{1/2,0}}  \; .  \label{fxyalt2}
\end{align}

Recall that $f_\mathbb H$ was chosen, so that it $f_{\cal R}$ would satisfy (\ref{prefactor}).  In the upper half plane,
\be
\langle \Phi_{1/2,0}(w, \bar w) \rangle_\mathbb{H}=\Imag\, w^{-2 h_{1/2,0}}\; ,
\ee
and our transformation into the rectangle gives
\be \label{1ptf}
\langle \Phi_{1/2,0}(z, \bar z)\rangle_{\cal R}=|w'(z)|^{2h_{1/2,0}}\langle \Phi_{1/2,0}(w(z), \bar w(z)) \rangle_\mathbb{H}=\left| \frac{\Imag\, w}{2K' \sqrt{w\left(m-w\right)\left(1-w\right)}} \right|^{-2 h_{1/2,0}}
\ee
comparing this to (\ref{fxy}) confirms (\ref{prefactor}).  In physical systems (\ref{1ptf}) can represents the density of clusters attached to a homogeneously wired boundary. To our knowledge no simple interpretation exists for $f_{\cal R}(z)$.

Finally, from (\ref{fxyalt1}), making use of (\ref{xipsidef}) and standard elliptic function properties, we find  (in a slight abuse of notation)
\be \label{fxipsi}
 f(\xi,\y,m) = c(m)  \left[ \frac{(1-m \xi^2)^2}{ \xi (1-\xi) (1-m \xi)} + \frac{(1-(1-m) \y^2)^2}{ \y (1-\y) (1-(1-m) \y)} -4 \right]^{-h_{1,3}/2+h_{1/2,0}} \; . 
\ee
This is the expression that we make use of in \cite{SimmonsKlebanFloresZiff11}, since it uses the mathematically natural coordinates $\xi$ and $\y$.

We note that the factor in brackets in (\ref{fxipsi}) is invariant under all three symmetry operations (\ref{Sym}).  Furthermore, the first two of these symmetries preserve $m$ and $f_{\cal R}(z)$ is completely invariant.

The third symmetry takes $m \to 1-m$. Combining (\ref{Rvsm}), (\ref{fconst}), and (\ref{fxipsi}) we see that this implies
\be
f(\xi,\y,m)=R^{-(4h_{1,2}^c+2 h_{1/2,0})}f\left(\y, \xi,1-m \right) \;.
\ee
The total scaling dimension of the operators in $C(z)$ is $4h_{1,2}^c+2 h_{1/2,0}$, and thus this additional factor is the conformal covariance factor of a uniform scaling by $1/R$. So $f_{\cal R}(z)$ is covariant, as required, under all symmetry operations (\ref{Sym}).

The factor $2^{h_{1,3}}$ in $c(m)$ is due to the corner operators.  In general, the corner operator convention introduces boundary-corner fusion rules 
\be
\phi_1(x+x_c) \phi_2^c(x_c)=C_{2;1}^{3\, c}x^{h_3^c-h_2^c-h_1} \phi_3^c(x_c)\; ,
\ee
which modify the regular boundary-boundary fusion rules when the expansion point is a corner $x_c$.  These corner OPE coefficients are related to the boundary OPE coefficients by $C_{2;1}^{3\, c}= 2^{h_1}C_{2;1}^{3}$, where $1,2$ and $3$ label the operators in the boundary OPE.  Thus the factor of $2^{h_{1,3}}$ in (\ref{fconst}) is included for consistency between the upper half-plane and rectangle expressions.

%%%%%%%%%%%%%%%%%%%%
%                  CONCLUSIONS                  %
%%%%%%%%%%%%%%%%%%%%
\section{Summary and discussion \label{Conc}}

\subsection{Summary}

We first summarize the main results.  They include  all solutions for the PDEs determining the six-point correlation function  (\ref{cf1})
$$ C(z)=\langle \phi_{1,2}^c(0)\phi_{1,2}^c(\I) \Phi_{1/2,0}(z, \bar z) \phi_{1,2}^c(R)\phi_{1,2}^c(R+\I) \rangle_{\cal R}\; ,$$ 
in the rectangle $\mathcal{R}:= \{z=x + \I y\in\mathbb{C}\, |\, 0<x<R, 0<y<1\}$. Making use of the coordinates $\xi(x,m)$ and $\y(y,m)$ from (\ref{xipsidef})
$$ \xi=\je{sn}{x}^2\; , \qquad \mathrm{and} \qquad \y=\mathrm{sn}\left(y\, K'|1-m\right)^2\; ,$$
with the elliptic parameter $m$ specifying the aspect ratio $R$ via (\ref{mvsR})
$$m=\frac{\vartheta_4{}^4\left(0,e^{- \pi R} \right)} {\vartheta_3{}^4\left(0,e^{- \pi R} \right)}\; ,$$
we find from (\ref{CfGeq}) that any solution which involves a single conformal block can be written in the form
$$C(z) = f(\xi,\y,m)G\left(\xi,m \right) \; ,$$ 
where the algebraic prefactor $f$ is given in (\ref{fxipsi})
$$ f(\xi,\y,m) = c(m)  \left[ \frac{(1-m \xi^2)^2}{ \xi (1-\xi) (1-m \xi)} + \frac{(1-(1-m) \y^2)^2}{ \y (1-\y) (1-(1-m) \y)} -4 \right]^{-\frac{1}{2}h_{1,3}+h_{1/2,0}} \; ,$$
with $c(m)$ from (\ref{fconst})
$$c(m) = 2^{ h_{1,3}}(K')^{8h_{1,2}+2 h_{1/2,0}}(m(1-m))^{2h_{1,2}} \; ,$$
and $G$ one of the ten solutions  (\ref{GI})--(\ref{GV}) or (\ref{GVI})--(\ref{GX}), or one of these solutions with $\{ \xi \rightarrow \y ,\, m   \rightarrow  1-m \}$. The prefactor $f$ is independent of the physical system, while $G$ is not.

A bit more technically, having chosen the algebraic prefactor $f_\mathbb{H}$ as in (\ref{UHP CI Form}), we employ standard methods of CFT to give  the PDEs for the conformal blocks.  Because of the $\phi_{1,2}$s in $C(z)$, these are second-order equations.  Introducing the coordinates $\xi$ and $\y$ (\ref{xipsidef}) then leads to an interesting independence from one coordinate  (\ref{DE3}), which may be written
$$\drv{\drv{F} \xi}\y(\xi, \y, m)=0 \; ,$$
and indicates some unknown symmetry.  As a consequence, conformal  blocks of $F$ only depend on two coordinates, not three. Some further manipulation transforms the PDEs into (\ref{AFDE}), which is the standard form for the Appell function $F_1$.    A solution set, as mentioned, is given in (\ref{GI})--(\ref{GV}) or (\ref{GVI})--(\ref{GX}).  The last five of these expressions are most simply expressed with the Appell function $F_2$. These solutions span a three-dimensional space; when the $\y$ sector is included it is six-dimensional. The relations between the solutions are given in (\ref{linrel1}) and (\ref{linrel2}).   Alternate forms for the algebraic prefactor $f$ in the rectangle are given in (\ref{fxy})--(\ref{fxyalt2}).

The results are of physical interest, since, for a variety of two-dimensional critical points, they specify the density of critical clusters  anchored to one or both opposite ends of a rectangle with wired boundary conditions on those ends.  These applications are discussed in \cite{SimmonsKlebanFloresZiff11}. 

Appendix \ref{F2Der} derives an Appell function relation that is useful in determining the $G$s, and \ref{Ccd} discusses  general conditions that must be satisfied for the conformal blocks, as here, to have a common $y$--dependence.

%%%%%%%%%%%%%%%%%%%%
%                 RELATED WORK                  %
%%%%%%%%%%%%%%%%%%%%
\subsection{Related work}

In the case $\k = 6$, recent work by Beliaev and Izyurov \cite{BeliaevIzyurov} gives a rigorous derivation of the PDEs  (\ref{DE1}) in the case $R \to \infty$ ($m \to 1$) using SLE techniques.  This provides a basis for a completely rigorous derivation of our results for the case of critical percolation on the triangular lattice in this limit.

\section{Acknowledgments}
We thank S. Flores for a careful reading of the manuscript.

This work was supported by EPSRC Grant No.\ EP/D070643/1 (JJHS), 
 and by the National Science Foundation Grants Nos. MRSEC DMR-0820054 (JJHS) and  DMR-0536927 (PK). %

%%%%%%%%%%%%%%%%%%%
%                  APPENDICES                  %
%%%%%%%%%%%%%%%%%%%
\appendix{}
%%%%%%%%%%%%%%%%%%%
%        APPELL FUNCTION                  %
%%%%%%%%%%%%%%%%%%%
\section{Appell Function relations} \label{F2Der}
In this section we derive a relation between Appell functions (\ref{Equ: Appell Identity}) used in deriving (\ref{GVI})--(\ref{GX}).  The manipulations involved are relatively simple, involving gamma functions, Pochhammer symbols and Gauss's hypergeometric function. We use standard identities (see \cite{AbSt} for example) and assume that $s,t \in [0,1)$. 

Beginning with the series expression for $F_1$ we rearrange the summation with $k=i+j$
\begin{align*}
F_1\left(a;b_1,b_2;c|s,s t\right)&=\sum_{k=0}^\infty\sum_{j=0}^k \frac{(a)_k(b_1)_{k-j}(b_2)_j s{}^k t{}^j}{(k-j)!\, j!\, (c)_{k}}\\
&=\sum_{k=0}^\infty  \frac{(a)_k (b_1)_k s{}^k}{k! (c)_{k}} \sum_{j=0}^k \frac{ (-k)_j (b_2)_jt{}^j}{(1-b_1-k)_j\, j!}
=\sum_{k=0}^\infty  \frac{(a)_k (b_1)_k s{}^k}{k! (c)_k} {}_2F_1(-k,b_2,1-b_1-k|t)\; ,
\end{align*}
where we used
\be\nn
\frac{(b_1)_{k-j}}{(k-j)!}=\frac{(b_1)_k (-k)_j}{ k!(1-b_1-k)_j}
\ee
which follows on rewriting the Pochhammer symbols, $(z)_n=\G(z+n)/\G(z)$ and using the identity
\be\nn
\G(z+n)\G(1-z-n)=(-1)^n\G(z)\G(1-z)=(-1)^n \pi \csc (z \pi)\;,
\ee
for $n\in \mathbb{Z}$.
We can change the argument from $t$ to $1-t$ using standard hypergeometric function identities with the result that
\begin{align*}
F_1\left(a;b_1,b_2;c|s,s t\right)&=\sum_{k=0}^\infty  \frac{(a)_k (b_1)_k s{}^k}{k! (c)_k}\frac{(1-b_1-b_2-k)_k}{(1-b_1-k)_k}{}_2F_1(-k,b_2,b_1+b_2|1-t)\\
&=\sum_{k=0}^\infty \sum_{j=0}^k  \frac{(a)_k (b_1+b_2)_k (-k)_j(b_2)_j s{}^k(1-t)^j}{k!j! (c)_k(b_1+b_2)_j}\\
&=\sum_{i=0}^\infty \sum_{j=0}^\infty  \frac{(a)_{i+j} (b_1+b_2)_{i+j} (-i-j)_j(b_2)_j s{}^{i+j}(1-t)^j}{(i+j)! j! (c)_{i+j} (b_1+b_2)_j}\\
&=\sum_{i=0}^\infty \sum_{j=0}^\infty  \frac{(a)_{i+j} (b_1+b_2)_{i+j} (b_2)_j s{}^{i+j}(t-1)^j}{i! j! (c)_{i+j} (b_1+b_2)_j}
\end{align*}
where the second and final expressions take advantage of the relation $(1-a-i)_i=(-1)^i(a)_i$.

Now we sum over the $i$ index first.  Here we use the identity $(a)_{i+j}=(a+j)_i(a)_j$, and define the quantity $\Delta:=c-b_1-b_2-a$ to simplify our expressions
\begin{align*}
F_1\left(a;b_1,b_2;c|s,s t\right)&=\sum_{j=0}^\infty  \frac{(a)_{j}(b_2)_j [s(t-1)]^j}{j! (c)_j}\sum_{i=0}^\infty \frac{(a+j)_{i} (b_1+b_2+j)_{i}s{}^i}{i!(c+j)_{i}}\\
&=\sum_{j=0}^\infty  \frac{(a)_{j}(b_2)_j [s(t-1)]^j}{j! (c)_j}{}_2F_1(a+j,b_1+b_2+j,c+j|s)\\
&=\sum_{j=0}^\infty  \frac{(a)_{j}(b_2)_j [s(t-1)]^j}{j! (c)_j}\bigg(
\frac{\G(-\Delta+j)\G(c+j)}{\G(a+j)\G(b_1+b_2+j)}(1-s)^{\Delta-j}{}_2F_1(c-a,\Delta+a,1+\Delta-j|1-s) \\
&\qquad +
\frac{\G(\Delta-j)\G(c+j)}{\G(c-a)\G(\Delta+a)}s^{1-c-j}{}_2F_1(1-\Delta-a,1+a-c,1-\Delta+j|1-s)
\bigg)\\
&=\frac{\G(c)\G(-\Delta)(1-s)^\Delta}{\G(a)\G(b_1+b_2)}\sum_{j=0}^\infty  \frac{(b_2)_j (-\Delta)_j}{j!(b_1+b_2)_j}\left[\frac{s(1-t)}{s-1}\right]^j{}_2F_1(c-a,\Delta+a,1+\Delta-j|1-s) \\
&\qquad +\frac{\G(c)\G(\Delta)s^{-a}}{\G(c-a)\G(\Delta+a)}\sum_{j=0}^\infty  \frac{(a)_{j}(b_2)_j (1-t)^j}{j!(1-\Delta)_j}{}_2F_1(a+j,1+a-c,1-\Delta+j|1-1/s)\\
\end{align*}
in the final two steps we use standard identities to change the argument from $s$ to $1-s$, and then in one case from $1-s$ to $1-1/s$.

If we write the functions in series form and manipulate the Pochhammer symbols we find
\begin{align} \nn
F_1\left(a;b_1,b_2;c|s,s t\right)&=\frac{\G(c)\G(-\Delta)(1-s)^\Delta}{\G(a)\G(b_1+b_2)}\sum_{i=0}^\infty \sum_{j=0}^\infty   \frac{(c-a)_i(\Delta+a)_i (b_2)_j (-\Delta)_j}{i!j!(b_1+b_2)_j(1+\Delta-j)_i} (1-s)^i \left[\frac{s(1-t)}{s-1}\right]^j \\ \nn
&\qquad +\frac{\G(c)\G(\Delta)s^{-a}}{\G(c-a)\G(\Delta+a)}\sum_{j=0}^\infty\sum_{i=0}^\infty  \frac{(a+j)_i(1+a-c)_i(a)_{j}(b_2)_j (1-1/s)^i(1-t)^j}{i!j!(1-\Delta)_{i+j}}\\ \nn
&=\frac{\G(c)\G(-\Delta)(1-s)^\Delta}{\G(a)\G(b_1+b_2)}\sum_{i=0}^\infty   \frac{(c-a)_i(\Delta+a)_i}{i!(1+\Delta)_i} (1-s)^i \sum_{j=0}^\infty \frac{ (b_2)_j (-\Delta-i)_j}{j!(b_1+b_2)_j} \left[\frac{s(1-t)}{s-1}\right]^j \\ \label{AppStep}
&\qquad +\frac{\G(c)\G(\Delta)s^{-a}}{\G(c-a)\G(\Delta+a)}F_1(a;1+a-c,b_2;1-\Delta| 1-1/s,1-t)
\end{align}

We'll need to consider $s,t\in(0,1)$, but neither of the terms in (\ref{AppStep}) converge for all values in this domain.  For the first term we perform the $j$ summation and use the identity
$$
{}_2F_1\left(b_2,-\Delta-i, b_1+b_2 \bigg| \frac{s(1-t)}{s-1} \right)=\left(\frac{1-s}{1-s t}\right)^{b_2}{}_2F_1\left(b_2,c-a+i, b_1+b_2 \bigg| \frac{s(1-t)}{1-s t} \right)\; ,
$$
replacing the problematic argument with one that gives  convergence in the desired range.  If we then explicitly write the new hypergeometric series we find the first term in (\ref{AppStep})  is
\begin{multline*}
\frac{\G(c)\G(-\Delta)(1-s)^{\Delta+b_2}}{\G(a)\G(b_1+b_2)(1-s t)^{b_2}}\sum_{i=0}^\infty\sum_{j=0}^\infty   \frac{(c-a)_{i+j}(\Delta+a)_i(b_2)_j}{i!j!(1+\Delta)_i(b_1+b_2)_j} (1-s)^i \left(\frac{s(1-t)}{1-s t}\right)^j\\
=\frac{\G(c)\G(-\Delta)(1-s)^{\Delta+b_2}}{\G(a)\G(b_1+b_2)(1-s t)^{b_2}}F_2\left(c-a;\Delta+a,b_2;1+\Delta,b_1+b_2\big|1-s,\frac{s(1-t)}{1-s t}\right)\; .
\end{multline*}
The $F_2$ series is convergent for arguments $x$ and $y$ such that $|x|+|y|<1$; it is straightforward to check that this expression is convergent for all $s,t\in(0,1)$.

Using known $F_1$ relations we can rewrite the Appell function from the second term of (\ref{AppStep}) as
$$
F_1(a;1+a-c,b_2;1-\Delta| 1-1/s,1-t)=s^{1+a-c}t^{1+b_1-c} F_1(1-\Delta-a;1+a-c,b_1;1-\Delta| 1-s t,1-t)
$$
which is convergent for our entire domain of $\{s,t\}$.

Putting these pieces together we find the identity
\begin{multline} \label{Equ: Appell Identity}
F_1\left(a;b_1,b_2;c|s,s t\right)=\frac{\G(c)\G(-\Delta)(1-s)^{\Delta+b_2}}{\G(a)\G(b_1+b_2)(1-s t)^{b_2}}F_2\left(c-a;\Delta+a,b_2;1+\Delta,b_1+b_2\bigg|1-s,\frac{s(1-t)}{1-s t}\right)\\
 +\frac{\G(c)\G(\Delta)t^{b_1}(st)^{1-c}}{\G(c-a)\G(\Delta+a)} F_1(1-\Delta-a;1+a-c,b_1;1-\Delta| 1-s t,1-t)\; .
\end{multline}

Using (\ref{Equ: Appell Identity}) to rewrite (\ref{GI}), we find $G_{\rm I}(\xi,m)=(G_{\rm II}(\xi,m)-G_{\rm VI}(\xi,m))/n$, with $G_{\rm VI}(\xi,m)$ as defined in (\ref{GVI}).  The relations for $G_{\rm VII}(\xi,m)$ through $G_{\rm X}(\xi,m)$ given in (\ref{GVI}--\ref{GX}) follow similarly.

%%%%%%%%%%%%%%%%%%%
%        Y-INDEPENDENCE                  %
%%%%%%%%%%%%%%%%%%%
\section{ Conditions for common y-dependence \label{Ccd} }

Finally we examine two related conditions for a set of correlation functions to exhibit common $y$ dependence in the rectangle $\mathcal{R}$.

One might wonder whether, when  $\Phi_{1/2,0}$ in (\ref{cf1}) is replaced by another bulk operator with a weight $2h$, the conformal blocks could still exhibit the $y$--independence (\ref{DE3}).  To answer this question we repeat the analysis in \ref{cfDEs} with all occurrences of $h_{1/2,0}$ changed to $h$.  We search for a solution $F(\xi,\y,m)=f_0(\xi,\y,m)F_0(\xi,\y,m)$ such that the relation $0=\drv{\drv{F_0}\xi} \y$ is implied by the differential equations.  The resulting conditions imply that $f_0$ must satisfy three differential equations, which only have a solution when $h=h_{1/2,0}$ and $f_0$ itself is constant.   In that sense our result is unique.   Note that the analysis depends on the second order differential equations, therefore the result is specific to the operators $\phi_{1,2}^c$. 

One can find a necessary condition in a somewhat different way that doesn't depend on the differential equations and could be adapted to other corner operators.  This method considers the weights of operators in the prospective correlation functions, along with their fusion rules.  We use the operators specific to the  correlation function (\ref{cf0}), but this condition is easily applied to any combination of operators that one might expect to exhibit this behavior and for which a similar understanding of the fusion rules exists.

Because the ratio of the correlation functions by hypothesis must not depend on the vertical position, the ratios of fusion products must be the same as $x \to 0$ whether we approach directly from the bulk ($0<y<1$) or first move to a horizontal side ($y=0$ or $1$).

The argument is most easily made by referring to the  $\mathrm{O}(n)$ loop gas.   Consider the limit $x \to0$, when the bulk point is taken to the left hand side of the rectangle. Then there are two cases; for one the leading order fusion product is the identity, for the other  a four leg operator, $\phi_{1,5}$. (This may be understood more explicitly via the treatment in \cite{SimmonsKlebanFloresZiff11}.) There are, correspondingly, two possible OPEs
\begin{align}
\phi_{1/2,0}(z, \bar z) &= C^{1/2,0}_{\mathbf{1}} x^{-2 h_{1/2,0}} \, \mathbf{1}\;, \qquad \mathrm{and} \\
\phi_{1/2,0}(z, \bar z) &= C^{1/2,0}_{\phi_{1,5}} x^{h_{1,5}-2 h_{1/2,0}} \, \phi_{1,5}(y)\; .
\end{align}
We assume that this result, and the similar one below, are generally valid and not restricted to $\mathrm{O}(n)$ models.

If, on the other hand, we first bring the density operator to the bottom edge it fuses to the two-leg operator $\phi_{1,3}(x)$. Only one operator can appear in this limit for the the conformal blocks.  If different operators appeared then the difference in scaling weights would eliminate the possibility of common $y$--dependence. We then move this operator into the lower corner of the rectangle, where there is a corner one-leg operator, $\phi^c_{1,2}$.  As mentioned in subsection \ref{Cff}, an operator in a corner of angle $\pi/2$ has a modified conformal weight, and thus $h^c= 2 h$.  Again, from the $\mathrm{O}(n)$ model there are two cases, either  the fusion product will be to a corner one-leg operator $\phi^c_{1,2}$ or the corner three-leg operator $\phi^c_{1,4}$.  The two possible OPEs are
\begin{align}
\phi_{1,3}(x) \phi^c_{1,2}(0) &= C^{1,3;1,2}_{1,2} x^{-h_{1,3}} \phi^c_{1,2}(0)\; \qquad \mathrm{and} \\
\phi_{1,3}(x) \phi^c_{1,2}(0) &= C^{1,3;1,2}_{1,4} x^{2 h_{1,4}-h_{1,3}-2 h_{1,2}} \phi^c_{1,4}(0)\; .
\end{align}

Thus, defining $\rho$ as the ratio of densities with $z$ not allowed to belong to the left end cluster to those with $z$ allowed to belong the the left end cluster, gives (depending on whether we approach via the bulk or the boundary)
\begin{align} \nonumber
\rho_{\mathrm{bulk}} &\sim x^{h_{1,5}}\\ \nonumber
\rho_{\mathrm{boundary}} &\sim x^{2(h_{1,4}-h_{1,2})}\; .
\end{align}
Since the ratio of two correlation functions doesn't depend on which approach we take, we must have
\be
h_{1,5} = 2(h_{1,4}-h_{1,2})\;
\ee
as a necessary condition for the factorization.

We can check this result with explicit expressions for the weights
\be
h_{1,2}=3/k-1/2\;, \qquad h_{1,4}=15/\k-3/2\;, \quad \mathrm{and}\quad h_{1,5}=24/\k-2\; ,
\ee
and we see that this condition is indeed true for arbitrary values of $\kappa$.

\bibliography{RedDensRZ}

\begin{thebibliography}{1}

\bibitem{BPZ84}
A.~A. {Belavin}, A.~M. {Polyakov}, and A.~B. {Zamolodchikov}.
\newblock {Infinite conformal symmetry in two-dimensional quantum field
  theory}.
\newblock {\em Nuclear Physics B}, 241:333--380, 1984.

\bibitem{BYB}
P.~{DiFrancesco}, P.~{Mathieu}, and D.~{S{\' e}n{\' e}chal}.
\newblock {\em Conformal Field Theory}.
\newblock Springer, 1999.

\bibitem{SimmonsZiffKleban08}
Jacob J.~H. Simmons, Robert~M. Ziff, and Peter Kleban.
\newblock Factorization of percolation density correlation functions for
  clusters touching the sides of a rectangle.
\newblock {\em J. Stat. Mech: Th. Exp.}, 2009:P02067, 2009.

\bibitem{SimmonsKlebanFloresZiff11}
Jacob J.~H. Simmons, Peter Kleban, Steven~M. Flores, and Robert~M. Ziff.
\newblock Cluster densities at 2-d critical points in rectangular geometries.
\newblock {\em http://www.arxiv.com/abs/1103.5691}, 2011.

\bibitem{ApKdF}
P.~{Appell} and M.~J. {Kamp{\' e} de F{\' e}riet}.
\newblock {\em Fonctions Hypergeometriques et Hyperspheriques}.
\newblock Gauthier-Villars, Paris, 1926.

\bibitem{Cardy84}
J.~L. Cardy.
\newblock Conformal invariance and surface critical behavior.
\newblock {\em Nucl. Phys.}, {\bf B240}:514--532, 1984.

\bibitem{BeliaevIzyurov}
D.~Beliaev and K.~Izyurov.
\newblock {A proof of a factorization formula for critical percolation.}
\newblock {\em arXiv:1011.5822v2}, 2010.

\bibitem{AbSt}
M.~Abramowitz and I.~A. Stegun.
\newblock {\em Handbook of Mathematical Functions}.
\newblock Dover Publications Inc., New York, 1964.

\end{thebibliography}
\end{document}